\documentclass[reprint,superscriptaddress,amssymb,amsmath,aps,showpacs,10pt,longbibliography,prl]{revtex4-2}
\usepackage{graphicx}
\pdfoutput=1

\graphicspath{{figures/}{../}}
\usepackage{color}
\usepackage[colorlinks,bookmarks=false,citecolor=darkblue,linkcolor=red,urlcolor=blue]{hyperref}
\def\cred{
}
\definecolor{darkred}{rgb}{0.7,0.0,0.0}
\definecolor{darkblue}{rgb}{0,0.02,0.45}

\def\cdbl{\color{darkblue}}
\definecolor{darkgreen}{rgb}{0.02,0.45,0.0}

\newcommand{\av}{\mathbf a}
\newcommand{\bv}{\mathbf b}
\newcommand{\cv}{\mathbf c}

\begin{document}


\title{Effect of nonhydrostatic pressure on the superconducting kagome metal CsV$_3$Sb$_5$}

\author{Alexander~A. Tsirlin}
\email{altsirlin@gmail.com}
\affiliation{Felix Bloch Institute for Solid-State Physics, Leipzig University, 04103 Leipzig, Germany}
\affiliation{Experimental Physics VI, Center for Electronic Correlations and Magnetism, Institute of Physics, University of Augsburg, 86135 Augsburg, Germany}

\author{Brenden R. Ortiz}
\affiliation{Materials Department, University of California Santa Barbara, Santa Barbara, CA, 93106, USA}
\affiliation{California Nanosystems Institute, University of California Santa Barbara, Santa Barbara, CA, 93106, USA}

\author{Martin Dressel}
\affiliation{1. Physikalisches Institut, Universit{\"a}t Stuttgart, 70569 Stuttgart, Germany}

\author{Stephen D. Wilson}
\affiliation{Materials Department, University of California Santa Barbara, Santa Barbara, CA, 93106, USA}

\author{Stephan Winnerl}
\affiliation{Helmholtz-Zentrum Dresden-Rossendorf, Institute of Ion Beam Physics \& Materials Research, D-01328 Dresden, Germany}

\author{Ece Uykur}
\email{e.uykur@hzdr.de}
\affiliation{1. Physikalisches Institut, Universit{\"a}t Stuttgart, 70569 Stuttgart, Germany}
\affiliation{Helmholtz-Zentrum Dresden-Rossendorf, Institute of Ion Beam Physics \& Materials Research, D-01328 Dresden, Germany}


\begin{abstract}
High-pressure single-crystal x-ray diffraction experiments reveal that the superconducting kagome metal CsV$_3$Sb$_5$ transforms from hexagonal ($P6/mmm$) to monoclinic ($C2/m$) symmetry above 10\,GPa if nonhydrostatic pressure conditions are created in a diamond anvil cell with silicon oil as the pressure-transmitting medium. This is contrary to the behavior of CsV$_3$Sb$_5$ under quasi-hydrostatic conditions in neon, with the hexagonal symmetry retained up to at least 20\,GPa. Monoclinic distortion leaves the kagome planes almost unchanged but deforms honeycomb nets of the Sb atoms. {\cred While the onset of the distortion almost coincides with the reentrance of superconductivity, our \textit{ab initio} density-functional calculations reveal only minor changes in the electronic structure compared to the quasi-hydrostatic case. In particular, Fermi surface reconstruction driven by the formation of interlayer Sb--Sb bonds is observed in both monoclinic and hexagonal CsV$_3$Sb$_5$ structures at high pressures and comes out as the likely cause for the reentrant behavior.}
\end{abstract}

\maketitle


{\cdbl\textit{Introduction.}} 
Envisaged theoretically a decade ago~\mbox{\cite{yu2012,kiesel2012,wang2013,kiesel2013}}, the superconductivity of kagome metals was recently realized in the family of $A$V$_3$Sb$_5$ compounds ($A$ = K, Rb, Cs) where vanadium atoms form an undistorted kagome lattice~\cite{ortiz2019,neupert2022}. The respective band structures strongly resemble those of a nearest-neighbor kagome metal and show van Hove singularities in the vicinity of the Fermi level~\cite{nakayama2021,cho2021,kang2022,hu2022}. These van Hove singularities, also known as band saddle points, may be responsible for two electronic instabilities observed in AV$_3$Sb$_5$, the superconductivity and charge-density wave (CDW)~\cite{neupert2022}. External stimuli are often used to tailor the electronic structure of materials and may be instrumental in reaching different microscopic regimes of kagome metals~\cite{qian2021,labollita2021,consiglio2022}. In this respect, the pressure-induced evolution of AV$_3$Sb$_5$ compounds and van Hove singularities in their electronic structures is of significant interest.

Here, we focus on CsV$_3$Sb$_5$ where the CDW transition is suppressed around 2\,GPa~\cite{chen2021a,yu2021}. The superconducting $T_c$ is enhanced from about 2.5\,K~\cite{ortiz2020} at ambient pressure to 8\,K at 2\,GPa~\cite{chen2021a,yu2021}. Further compression leads to a gradual reduction in $T_c$ followed by its re-entrant behavior above $12-15$\,GPa, with the second superconducting dome that persists well above 40\,GPa and reaches $T_c$ of about 5\,K~\cite{zhang2021a,chen2021b,yu2022}. X-ray diffraction (XRD) study of single crystals reveals that the [V$_3$Sb$_5$] slabs, which are well separated at ambient pressure, become increasingly closer to each other upon compression, resulting in the formation of interlayer Sb--Sb bonds~\cite{tsirlin2022}. The corresponding reconstruction of the Fermi surface at the pressure of about 12\,GPa correlates with the re-entrant behavior of superconductivity.

The information on superconductivity and pressure-dependent $T_c$ is obtained from transport measurements that are typically performed with silicon oil or Daphne oil as the pressure-transmitting medium~\cite{yu2021,chen2021a}. At higher pressures, these oils lead to increasingly nonhydrostatic conditions~\cite{klotz2009,perez2011}.
While quasi-hydrostatic conditions were realized in the previous single-crystal XRD experiment with neon gas as pressure-transmitting medium~\cite{tsirlin2022}, XRD experiments on powders~\cite{zhang2021a,yu2022} were performed with oil but did not allow a complete structure refinement. It remains unknown whether deviations from hydrostaticity, which necessarily occur in transport measurements, affect the CsV$_3$Sb$_5$ crystal structure and in what way.

In the following, we uncover the impact of nonhydrostaticity on CsV$_3$Sb$_5$. Using silicon oil as the pressure-transmitting medium in the single-crystal XRD experiment, we identify a monoclinic distortion of the CsV$_3$Sb$_5$ crystal structure under nonhydrostatic pressure conditions
{\cred The distortion sets in above 10\,GPa and correlates with the reentrant superconductivity of CsV$_3$Sb$_5$. Previous studies suggested that similar deviations from the hexagonal symmetry may cause the reentrant behavior in pressurized kagome metals~\cite{du2022,shi2022}. Here, we fully resolve the distorted structure for the first time and show that only weak changes in the electronic structure occur as a result of the monoclinic distortion. In contrast, the effect of Fermi surface reconstruction due to the formation of interlayer Sb--Sb bonds is generic for both hexagonal and monoclinic phases. This effect is also concomitant with the reentrant behavior and should be central to the pressure evolution of CsV$_3$Sb$_5$.
}
\smallskip


{\cdbl\textit{Methods.}} XRD measurements were performed on a $50\times 30\times 10$\,$\mu$m single crystal of CsV$_3$Sb$_5$ from the same batch as in Ref.~\cite{tsirlin2022}. The experiment was done in the same diamond anvil cell with the 500\,$\mu$m culet size and using the same setup of the CRYSTAL beamline at the SOLEIL synchrotron (France). Only the pressure-transmitting medium was changed to silicon oil, whereas in Ref.~\cite{tsirlin2022} neon gas was used. Deviations from hydrostaticity have already been reported in silicon oil at $4-5$\,GPa~\cite{klotz2009,perez2011}, whereas neon is expected to provide quasi-hydrostatic conditions up to at least 15\,GPa~\cite{klotz2009,takemura2021}. {\cred Pressure was determined by measuring the lattice parameter of the gold powder~\cite{supplement} used as the internal standard with the calibration from Ref.~\cite{fei2007}.}

The x-ray wavelength was 0.42438\,\r A. Diffraction data were collected by $1^{\circ}$ $\varphi$-rotations of the diamond anvil cell. \texttt{Crysalis Pro} software~\cite{crysalispro} was used for integration and absorption correction. \texttt{Jana2006} was used for the structure refinement~\cite{jana2006}. {\cred Samples of the data and examples of the structure refinement are shown in the Supplemental Material~\cite{supplement}.} All experiments were performed at room temperature and thus characterize the normal state of CsV$_3$Sb$_5$, which is a prerequisite for electronic instabilities at low temperatures.

\begin{figure}
\includegraphics{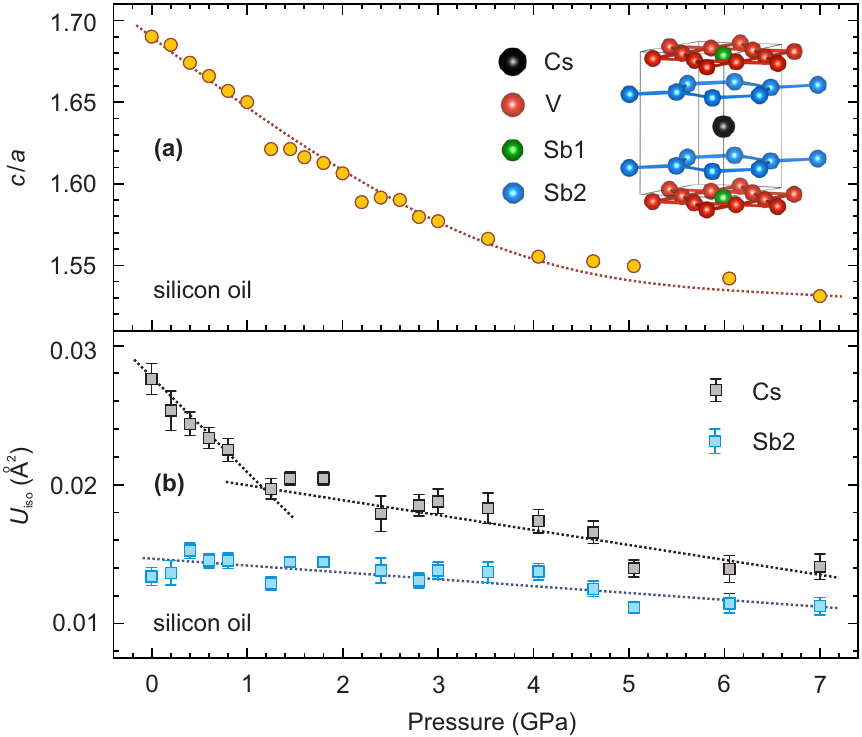}
\caption{\label{fig:lowp}
Pressure evolution of CsV$_3$Sb$_5$ crystal structure derived from XRD experiments with silicon oil in the hydrostatic regime: the $c/a$ ratio (a), and atomic displacement parameters $U_{\rm iso}$ (b). The lines are guide-for-the-eye only. The inset in (a) shows the crystal structure built by kagome planes of V and honeycomb planes of Sb2. 
}
\end{figure}

Density-functional band-structure calculations were performed in the \texttt{FPLO} code~\cite{fplo} using experimental lattice parameters and atomic positions. The Perdew-Burke-Ernzerhof (PBE) flavor of the exchange-correlation potential~\cite{pbe96} was used. Band structures were converged on the $k$ mesh with $24\times 24\times 12$ points within the first Brillouin zone. Additionally, we used \texttt{VASP}~\cite{vasp1,vasp2} to optimize the crystal structure of CsV$_3$Sb$_5$ at a constant volume and compare the experimental compressibility of CsV$_3$Sb$_5$ with \textit{ab initio} results obtained using different approximations for the exchange-correlation potential.
\smallskip


{\cdbl\textit{Hexagonal structure.}} Contrary to high-pressure XRD experiments on powder samples that are often plagued by an extensive reflection overlap and strong preferred orientation, measurements on single crystals allow a complete structure refinement. We first report the results obtained at pressures below 5\,GPa where hydrostatic conditions are obtained even with silicon oil. The results are in a good agreement with the earlier data~\cite{tsirlin2022} and show the rapid shrinkage of the crystal along $c$. No deviations from the hexagonal symmetry ($P6/mmm$) were observed, and diffraction images could be integrated with the low $R_{\rm int}$ of 0.03--0.04~\footnote{The $R_{\rm int}$ parameter shows the quality of averaging across same reflections measured on different images and across those reflections that are equivalent by symmetry. Higher values of $R_{\rm int}$ indicate that diffraction images are not consistent with the chosen structural symmetry.}. Intensities of about 370 reflections with 91 unique reflections were used to refine 5 structural parameters ($z$-coordinate of Sb2 as well as $U_{\rm iso}$ of Cs, V, Sb1, and Sb2)~\cite{supplement}.

\begin{figure}
\includegraphics{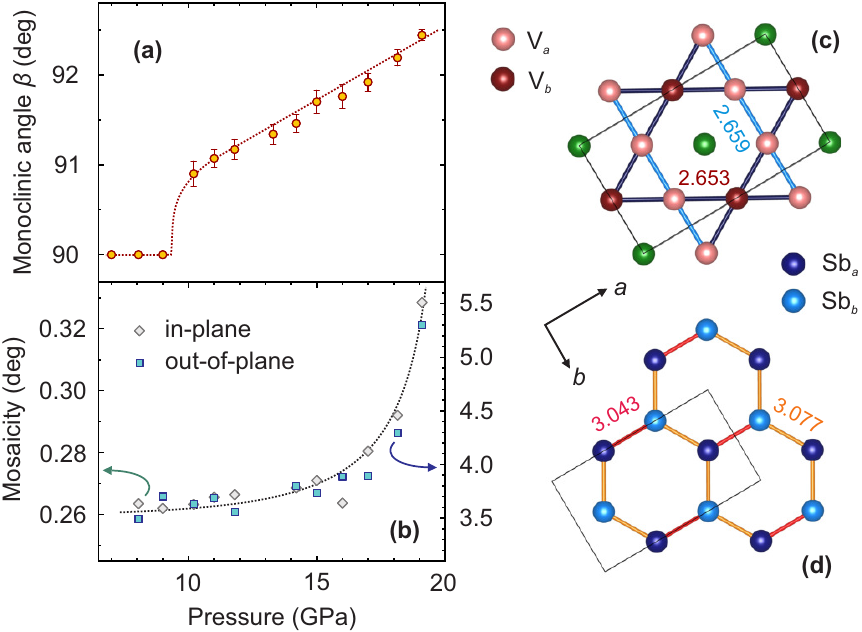}
\caption{\label{fig:monoclinic}
Pressure evolution of the monoclinic angle $\beta$ (a) and crystal mosaicity (b). The lines are guide-for-the-eye only. Deformation of the kagome planes (c) and honeycomb nets (d), with the representative interatomic distances (given in~\r A) obtained from the structure refinement at 19.1\,GPa.
}
\end{figure} 

The most remarkable result in this pressure range is the evolution of the atomic displacement parameter (ADP, $U_{\rm iso}$) for Cs, Fig.~\ref{fig:lowp}b. It is the only ADP that shows a significant pressure dependence. The high ADP at ambient pressure indicates the rattling motion of Cs atoms that are weakly bound between the [V$_3$Sb$_5$] slabs. The shrinkage of the structure along $c$ leads to a systematic reduction in $U_{\rm iso}$, with a notable kink around 1\,GPa that has no counterpart in the pressure dependence of the lattice parameters (Fig.~\ref{fig:lowp}a). It may indicate that some of the soft phonons abruptly become harder around this pressure. The kink in $U_{\rm iso}$ correlates with the non-monotonic evolution of $T_c$ showing a peak around 1\,GPa before reaching another peak at 2\,GPa~\cite{yu2021,chen2021a} (on a complete suppression of the CDW~\cite{wang2022}) and with recent conjectures on changes in the type of CDW at $0.5-1.0$\,GPa~\cite{li2022,gupta2022}.
\smallskip

\begin{figure}
\includegraphics{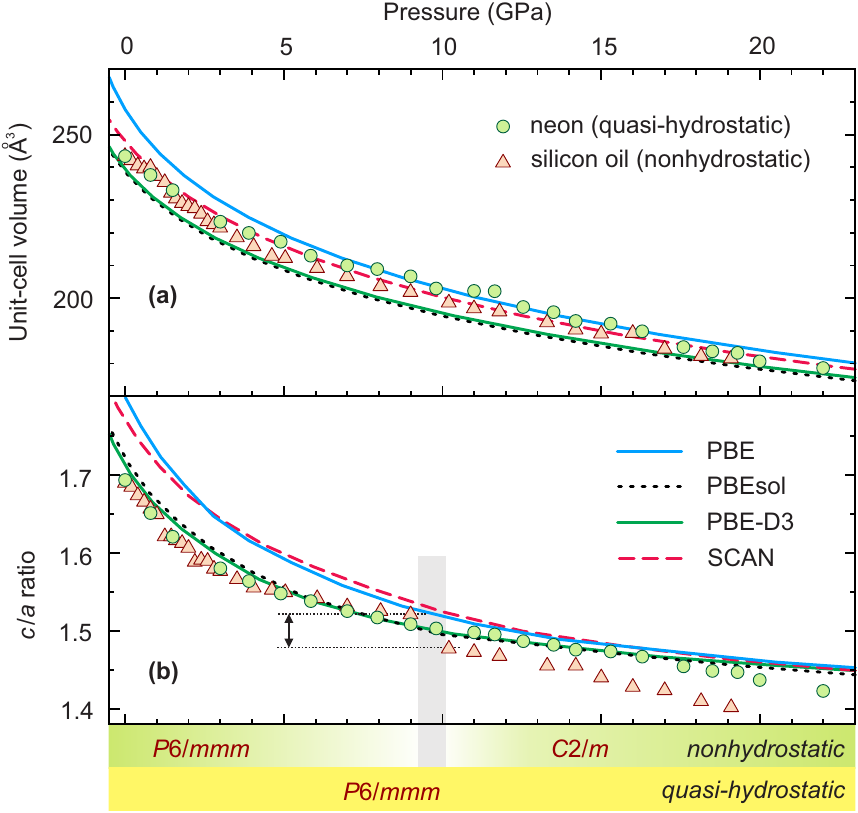}
\caption{\label{fig:compressibility}
Experimental unit-cell volume (a) and $c/a$ ratio (b) for CsV$_3$Sb$_5$ measured under both quasi-hydrostatic~\cite{tsirlin2022} and nonhydrostatic conditions. The lines show predictions obtained with different DFT functionals, as explained in the text. Note the abrupt change in $c/a$ around 10\,GPa when measured under nonhydrostatic conditions. The $c/a$ ratio of the monoclinic phase is estimated as $(c\sin\beta)/\bar a$ with $\bar a=(a/\sqrt{3}+b)/2$.%
}
\end{figure}

{\cdbl\textit{Monoclinic distortion.}} The hexagonal structural model perfectly describes the data up to 10\,GPa. At higher pressures, diffraction images could still be integrated assuming hexagonal symmetry, but with $R_{\rm int}$ rapidly increasing to $0.1-0.2$ and eventually making refinement of the hexagonal model impossible. This was not the case in our previous experiments under quasi-hydrostatic conditions~\cite{tsirlin2022} where even at 20\,GPa hexagonal symmetry allowed an integration of the data with $R_{\rm int}$ of about 0.06 and the structure refinement with the same residuals as at low pressures. In contrast, diffraction images obtained under nonhydrostatic conditions above 10\,GPa are better integrated in the monoclinic symmetry with the typical $R_{\rm int}$ of $0.04-0.06$. The continuous evolution of the unit-cell volume (Fig.~\ref{fig:compressibility}a) and the gradual departure of the monoclinic angle $\beta$ from $90^{\circ}$ (Fig.~\ref{fig:monoclinic}a) both suggest a second-order nature of the transition that implies the group-subgroup relation between the low-pressure and high-pressure polymorphs. Indeed, extinction conditions ($hkl$, $h+k=2n$) favored a $C$-centered monoclinic space group, which is obtained by the $\av_m=\av_h+\bv_h$, $\bv_m=\av_h-\bv_h$, $\cv_m=\cv_h$ lattice transformation. The eventual structure refinement was performed in $C2/m$, which is the subgroup of $P6/mmm$. 

\begin{table}
\caption{\label{tab:monoclinic}
Crystal structure of monoclinic CsV$_3$Sb$_5$ at 14.2\,GPa. The space group is $C2/m$. The lattice parameters are $a=9.199(2)$\,\r A, $b=5.341(2)$\,\r A, $c=7.75(3)$\,\r A, and $\beta=91.46(10)^{\circ}$. 
}
\begin{ruledtabular}
\begin{tabular}{cccccc}\smallskip
         & site & $x/a$ & $y/b$   & $z/c$    & $U_{\rm iso}$ (\r A$^2$) \\
 Cs      & $2a$ &   0   &   0     &   0      &   0.011(1)               \\
 V$_a$   & $4f$ &  0.25 &  0.75   &  0.5     &   0.007(2)               \\
 V$_b$   & $2d$ &  0.5  &   0     &  0.5     &   0.008(2)               \\
 Sb1     & $2c$ &  0.5  &  0.5    &  0.5     &   0.009(1)               \\
 Sb2$_a$ & $4i$ & 0.3308(4) & 0   & 0.221(1) &   0.011(1)               \\
 Sb2$_b$ & $4i$ & 0.1634(4) & 0.5 & 0.219(1) &   0.011(1)               \\
\end{tabular}
\end{ruledtabular}
\end{table}

Intensities of 237 reflections (130 unique) were used to refine 10 parameters of the monoclinic structure (Table~\ref{tab:monoclinic})~\cite{supplement}. Its main difference from the hexagonal one is the splitting of both V and Sb2 sites into two each. This split allows a deformation of both V kagome planes and Sb2 honeycomb nets, the two subunits of the [V$_3$Sb$_5$] slabs (inset of Fig.~\ref{fig:lowp}a), but only the honeycomb nets develop a distortion, with the Sb2--Sb2 distances of, respectively, 3.043(9)\,\r A and 3.077(5)\,\r A at 19.1\,GPa. By contrast, the nearest-neighbor V--V distances of, respectively, 2.6588(6)\,\r A and 2.6526(4)\,\r A at the same pressure suggest the absence of deformation in the kagome planes. Both V kagome planes and Sb2 honeycomb nets remain flat. The interlayer Sb2--Sb2 distance of 3.22(3)\,\r A at 19.1\,GPa is shorter than 3.35(2)\,\r A obtained under hydrostatic conditions at the same pressure. This happens because of the stronger shrinkage of the crystal structure along $c$. 

Above 15\,GPa, the increasing mosaicity (averaged reflection width) indicated a gradual crystal decay (Fig.~\ref{fig:monoclinic}) that coincides with the rapidly increasing deviations from hydrostaticity in the silicon oil~\cite{klotz2009}. On reaching the maximum pressure of 19.1\,GPa, the crystal was slowly decompressed. Measurements taken on decompression indicated that the hexagonal-monoclinic transition was reversible~\cite{supplement}. The data collected on decompression below 4\,GPa showed hexagonal symmetry with almost the same lattice parameters as on compression. However, the two-fold increase in the mosaicity indicated a significant crystal damage caused by the nonhydrostatic compression and led to unstable structure refinements for the data collected on decompression. 
\smallskip


{\cdbl\textit{Compressibility.}} Crystallographic parameters of pressurized CsV$_3$Sb$_5$ are often obtained from DFT-based crystal structure optimization that leads to different results depending on the choice of the exchange-correlation potential. Here, we benchmark DFT results for the compressibility of CsV$_3$Sb$_5$ using our experimental data. Fig.~\ref{fig:compressibility} and Table~\ref{tab:elastic} show that the most common PBE functional~\cite{pbe96} strongly overestimates the equilibrium volume of CsV$_3$Sb$_5$, but gradually improves at higher pressures where the interlayer bonding becomes stronger because of the Sb2--Sb2 bonds~\cite{tsirlin2022}. The strongly constrained and appropriately normed (SCAN) functional~\cite{scan} largely remedies the problem with the equilibrium volume and reproduces the unit-cell volume across the whole pressure range but overestimates $c/a$ similar to PBE. The best agreement for $c/a$ is reached by adding Grimme's D3 correction~\cite{grimme2010} to PBE or by using the PBEsol functional~\cite{pbesol} optimized for solids. It probably means that even at low pressures the interlayer interactions are not dispersive in nature, although they can be successfully captured by PBE+D3, which is typically used for layered systems such as van der Waals crystals.

\begin{table}
\caption{\label{tab:elastic}
Experimental and computed compression parameters of CsV$_3$Sb$_5$. $V_0$ stands for the equilibrium volume. The bulk modulus $B_0$ and its pressure derivative $B_0'$ are obtained by fitting the pressure dependence of unit-cell volume (experiment~\cite{tsirlin2022}) and volume dependence of energy (DFT) with the second-order Birch-Murnaghan equation of state.
}
\begin{ruledtabular}
\begin{tabular}{ccccc}\smallskip
            & $V_0$ (\r A$^3$/f.u.) & $c/a(P=0)$ & $B_0$ (GPa) & $B_0'$ \\
 Experiment & 243.41(3)             & 1.6940(1)  &  39(2)      & 3.9(3) \\
 PBE        & 257.8(4)              &  1.801(1)  &  16.1(6)    & 8.6(3) \\
 PBEsol     & 238.7(2)              &  1.723(1)  &  20.4(4)    & 9.4(2) \\
 PBE+D3     & 239.4(1)              &  1.715(1)  &  20.5(3)    & 9.5(2) \\
 SCAN       & 248.0(1)              &  1.768(1)  &  21.6(3)    & 7.6(1) \\
\end{tabular}
\end{ruledtabular}
\end{table}

It is also worth noting that none of the functionals allows an accurate description of both unit-cell volume and $c/a$ at the same time. This shortcoming becomes especially clear when the bulk modulus and its pressure derivative are compared to the experiment (Table~\ref{tab:elastic}). Therefore, experimental structural parameters obtained by single-crystal XRD remain the best choice for calculating the electronic structure of CsV$_3$Sb$_5$ under pressure.
\smallskip

\begin{figure*}
\includegraphics{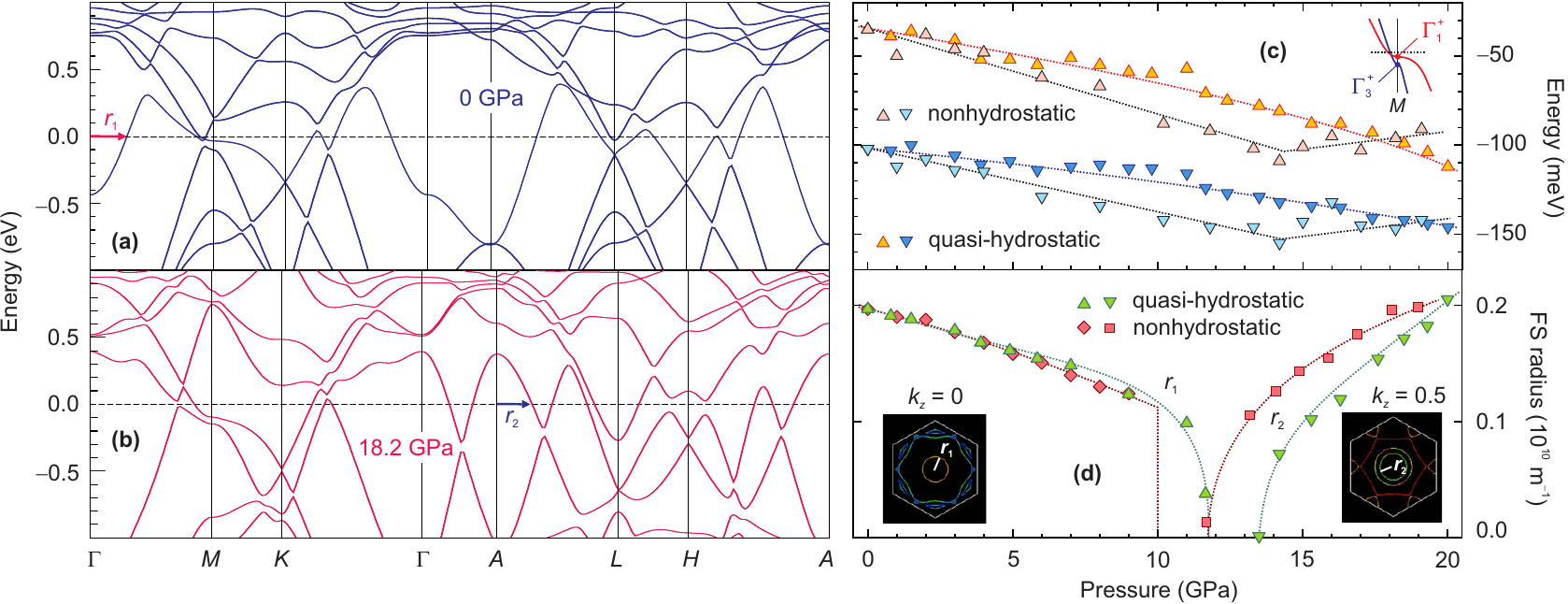}
\caption{\label{fig:bands}
Band structure of CsV$_3$Sb$_5$ at 0\,GPa (hexagonal, a) and 18.2\,GPa (monoclinic, b) displayed within the Brillouin zone of the hexagonal crystal structure. Pressure dependence of the energies of band saddle points at $M$ / van Hove singularities (c) and radii of the Sb $5p$ Fermi surface at $k_z=0$ and 0.5 (d). The insets in (d) show cross-sections of the relevant Fermi surfaces. All dotted lines are guide-for-the-eye only.
}
\end{figure*}

{\cdbl\textit{Electronic structure.}} 
The Fermi surface of CsV$_3$Sb$_5$ features contributions from both V $3d$ and Sb $5p$ bands~\cite{tsirlin2022}. Vanadium bands produce van Hove singularities that lie slightly below the Fermi level and may be responsible for exotic properties of AV$_3$Sb$_5$~\cite{neupert2022}. We track the positions of these van Hove singularities using band saddle points at $M$ (Fig.~\ref{fig:bands}). Both quasi-hydrostatic and nonhydrostatic conditions lead to a shift of the van Hove singularities away from the Fermi level. While this shift is monotonic in the former case, in agreement with the previous \textit{ab initio} studies~\cite{labollita2021,consiglio2022}, it is initially faster under nonhydrostatic conditions before the positions of the van Hove singularities saturate and even slightly move upwards above 13\,GPa (Fig.~\ref{fig:bands}c). These positions are controlled by the V $3d$ bandwidth that increases upon compression following the shortening of the lattice parameter $a$. Fig.~\ref{fig:compressibility}b illustrates that at 10\,GPa the $c/a$ ratio is higher in the nonhydrostatic case compared to the quasi-hydrostatic one. Therefore, the $a$ parameter shrinks faster, and the shift of the van Hove singularities away from the Fermi level is more pronounced. On the other hand, the monoclinic phase features a lower $c/a$, the $a$ parameter shrinks only weakly, and the positions of the van Hove singularities do not change or even slightly move back toward the Fermi level.

Similar arguments explain the evolution of Sb $5p$ bands around the Fermi level. Here, the main effect is the shrinkage and eventual disappearance of the Fermi surface pocket around $\Gamma$, while at higher pressures a new Fermi surface pocket appears around $A$ (Fig.~\ref{fig:bands}a,b) because covalent bonding between the Sb2 atoms in the adjacent honeycomb nets builds up~\cite{tsirlin2022}. In the nonhydrostatic case, this Fermi surface reconstruction happens at lower pressures (Fig.~\ref{fig:bands}d). The transition to the monoclinic structure eliminates the Fermi surface pocket at $\Gamma$ because monoclinic distortion results in the lower $c/a$ and, consequently, in the shorter Sb--Sb distances along the $c$ direction. Regardless of the pressure conditions, the Fermi surface reconstruction takes place in about the same pressure range of $10-15$\,GPa where reentrant behavior of the superconductivity has been observed~\cite{chen2021b,zhang2021a,yu2022}.%
\smallskip


{\cdbl\textit{Discussion.}} 
{\cred 
Reentrant superconductivity in CsV$_3$Sb$_5$ and other kagome metals remains vividly debated. From a theoretical standpoint, the unusual properties of these materials should be driven by van Hove singularities in the vicinity of the Fermi level~\cite{neupert2022,wu2021,park2021,ding2022,tazai2022,lin2022}. However, pressure -- both hydrostatic and nonhydrostastic -- systematically shifts the singularities away from the Fermi level (Fig.~\ref{fig:bands}c), so they are unlikely to be responsible for the reentrant behavior. Several studies pointed out structural phase transitions that are concomitant with the (re)appearance of superconductivity in pressurized kagome metals~\cite{du2022,shi2022}. However, no pressure-induced structural phase transition has been reported in CsV$_3$Sb$_5$ so far~\cite{zhang2021a,yu2022}. Here, we showed that such a transition does happen under nonhydrostatic pressure. We further resolved the structural distortion and elucidated its impact on the electronic structure. Since this impact is weak, the structural phase transition is unlikely to be responsible for the reentrant behavior, at least in CsV$_3$Sb$_5$, even though it accompanies the onset of the second superconducting dome. We also show that, regardless of pressure conditions, the interlayer Sb--Sb bonds form upon compression and cause a Fermi surface reconstruction. It is the most significant change in the electronic structure at $10-15$\,GPa and, thus, the likely origin of the reentrant behavior.
}

{\cred
Our data further suggest that the nonhydrostatic behavior of the silicon oil leads to a rapid degradation of the CsV$_3$Sb$_5$ crystal above 15\,GPa (Fig.~\ref{fig:monoclinic}b). The conditions of our XRD measurement, where one single crystal of CsV$_3$Sb$_5$ is immersed into silicon oil, are strongly resemblant to transport measurements in the diamond anvil cell, in contrast to XRD measurements on powder where small grains of the sample are mixed with silicon oil. Although powder XRD data on CsV$_3$Sb$_5$ still show diffraction peaks well above 20\,GPa~\cite{zhang2021a,yu2022}, it is likely that crystals undergo significant damage in this pressure range, and superconductivity is measured on a granular sample. This explains discrepancies in the published data for the second superconducting dome, especially the difference in the maximum $T_c$ values reported therein~\cite{chen2021b,zhu2022}. More generally, we show that hydrostaticity can have a strong influence on the structural symmetry of kagome metals and needs to be taken into consideration when pressure-induced behavior is analyzed.%
}

\smallskip


{\cdbl\textit{Conclusions.}} 
We have shown that the symmetry of CsV$_3$Sb$_5$ in its normal state is reduced when pressure conditions are nonhydrostatic. The monoclinic distortion observed above 10\,GPa leaves the kagome planes almost intact but deforms the Sb2 honeycomb nets and affects the $c/a$ ratio. These features do not change the main trends in the pressure evolution of the electronic structure, yet they facilitate the Fermi surface reconstruction by reducing the interlayer Sb--Sb distance compared to the quasi-hydrostatic case. {\cred Our results suggest that at least in CsV$_3$Sb$_5$, the structural phase transition is unlikely to be responsible for the reentrant behavior. In contrast, the evolution of the Sb bands and the ensuing reconstruction of the Fermi surface seem to play a major role.} More generally, we show that nonhydrostatic conditions incurred in high-pressure transport measurements can have repercussions for the crystal structure and even change its symmetry. The effect of nonhydrostaticity should be carefully evaluated when pressure is used as the tuning knob for exotic metals and their electronic properties.

\acknowledgments
We are grateful to Gabriele Untereiner for preparing single crystals for the XRD experiment. We thank SOLEIL for providing the beamtime and Pierre Fertey for his technical support during the measurement. S.D.W. and B.R.O. gratefully acknowledge support via the UC Santa Barbara NSF Quantum Foundry funded via the Q-AMASE-i program under award DMR-1906325. B.R.O. also acknowledges support from the California NanoSystems Institute through the Elings fellowship program. The work has been supported by the Deutsche Forschungsgemeinschaft (DFG) via Grants No. DR228/51-1 and UY63/2-1. E.U. acknowledges the European Social Fund and the Baden-W\"urttemberg Stiftung for the financial support of this research project by the Eliteprogramme. Computations for this work
were done (in part) using resources of the Leipzig University Computing Center.


%

\clearpage\newpage
\begin{widetext}
\begin{center}
\large\textbf{\textit{Supplemental Material}\smallskip \\ Effect of nonhydrostatic pressure on the superconducting kagome metal CsV$_3$Sb$_5$}
\end{center}

\renewcommand{\thefigure}{S\arabic{figure}}
\renewcommand{\thetable}{S\arabic{table}}
\setcounter{figure}{0}

Here, we summarize technical details of the structure determination by single-crystal XRD:
\medskip

1) Lattice parameters and atomic positions for CsV$_3$Sb$_5$ as a function of pressure (Fig.~\ref{fig:parameters})
\smallskip

2) Sample structure refinements that show a good match between the experimental and calculated reflection intensities, as well as the $R$-values at different pressures (Fig.~\ref{fig:refinement})
\smallskip

3) Reciprocal-lattice reconstructions that illustrate deviations from the hexagonal symmetry in the monoclinic phase and the restored hexagonal symmetry on decompression (Fig.~\ref{fig:reclattice})
\smallskip

4) Evolution of gold reference during the experiment with silicon oil (Fig.~\ref{fig:gold}). Here, the increasing stress leads not only to a gradual reflection broadening, but also to dissimilar shifts of the reflections, such that they can no longer be indexed in the cubic unit cell. We used angular position of the 222 reflection to determine pressure according to the calibration from Ref.~\cite{fei2007}. Above 12\,GPa, the resulting unit-cell volume of gold deviates from the literature data measured under quasi-hydrostatic conditions~\cite{takemura2001}.
\medskip

We also show a direct comparison of the band structures calculated for the hexagonal and monoclinic phases of CsV$_3$Sb$_5$ at the same pressure (Fig.~\ref{fig:bands-SM}). 
\vspace{0.7cm}

\begin{figure}[!h]
\includegraphics[width=12cm]{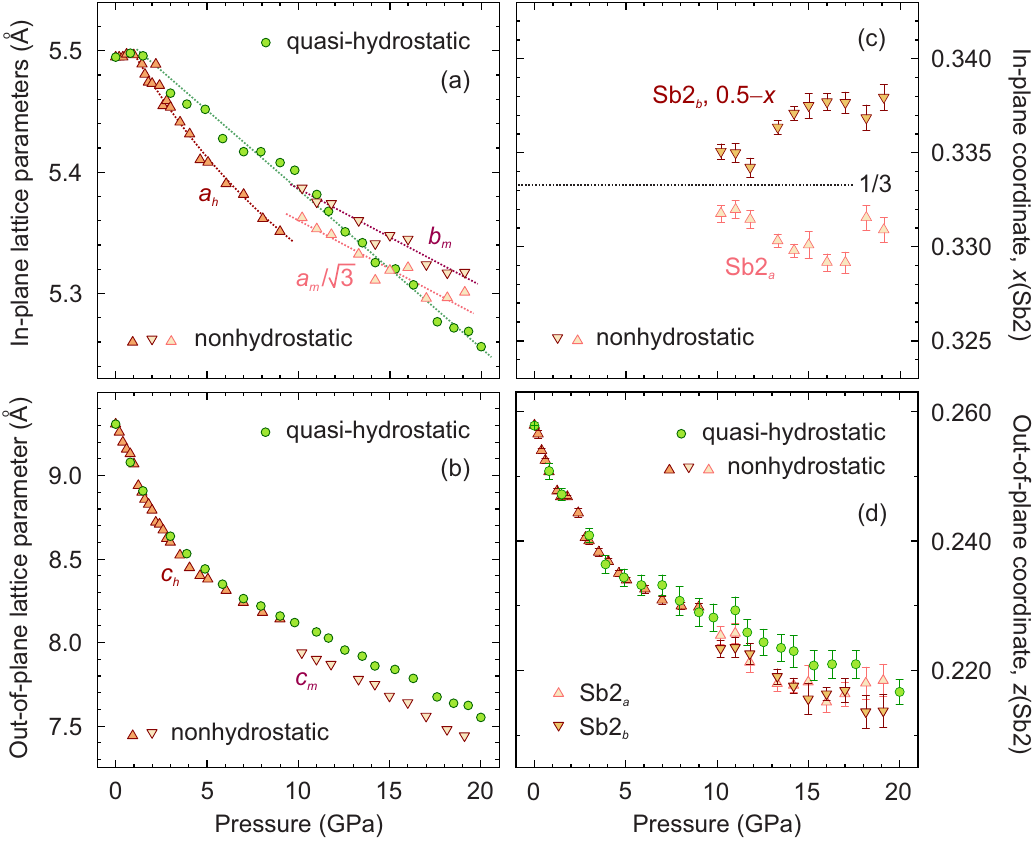}
\caption{\label{fig:parameters}
Pressure dependence of the lattice parameters (a,b) and Sb2 coordinates (c,d). The data for quasi-hydrostatic conditions are from Ref.~\cite{tsirlin2022}. The $x$-coordinate of Sb2 is $\frac13$ in the hexagonal structure. The lines are guide-for-the-eye only.
}
\end{figure}

\begin{figure}
\includegraphics[width=16cm]{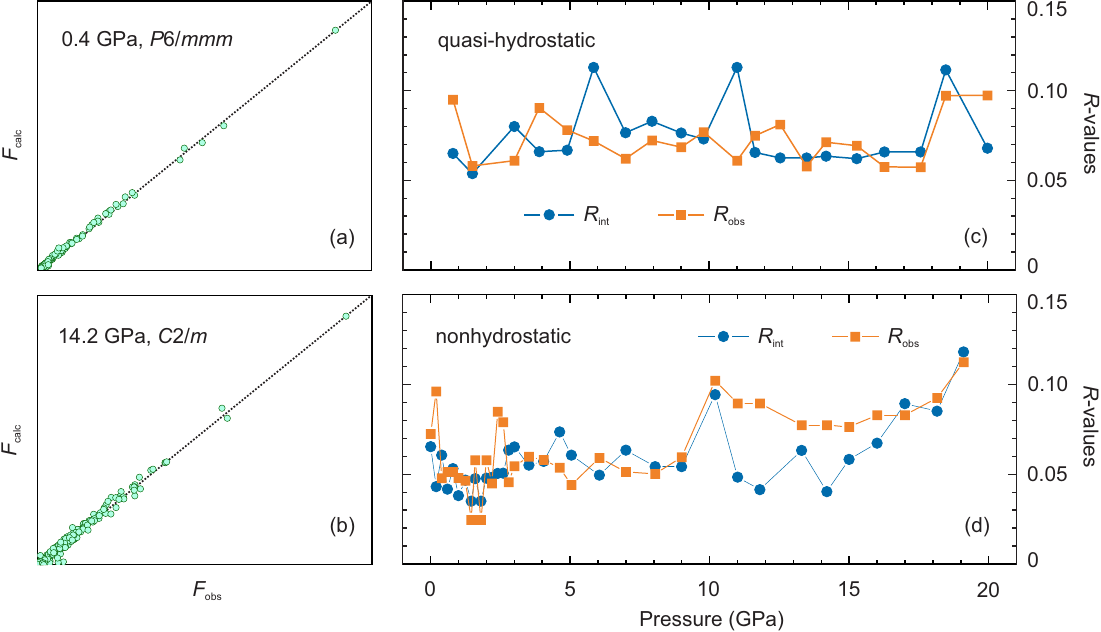}
\caption{\label{fig:refinement}
(a,b) Comparison between the observed and calculated reflection intensities for the data collected under nonhydrostatic conditions at 0.4\,GPa (hexagonal phase) and 14.2\,GPa (monoclinic phase). (c,d) Refinement residuals at different pressures. The results for the quasi-hydrostatic conditions are from Ref.~\cite{tsirlin2022}. Whereas the $R_{\rm int}$ values show the quality of averaging over equivalent reflections, the values of $R_{\rm obs}$ illustrate the match between the observed and calculated intensities (quality of the structure refinement). The monoclinic $C2/m$ model is used for the data measured under nonhydrostatic conditions above 10\,GPa.
}
\end{figure}

\begin{figure}
\includegraphics[width=15cm]{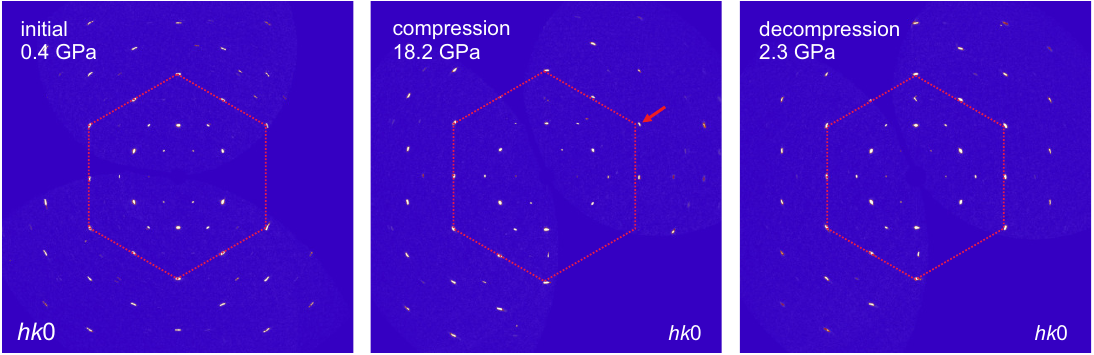}
\caption{\label{fig:reclattice}
Reciprocal-lattice reconstruction in the $hk0$ plane shows deviation from the hexagonal symmetry upon compression to 18.2\,GPa (the reflections do not lie on the regular hexagon, especially the reflection shown by the arrow) and revival of the hexagonal symmetry upon further decompression to 2.3\,GPa. 
}
\end{figure}

\begin{figure}
\includegraphics[width=12cm]{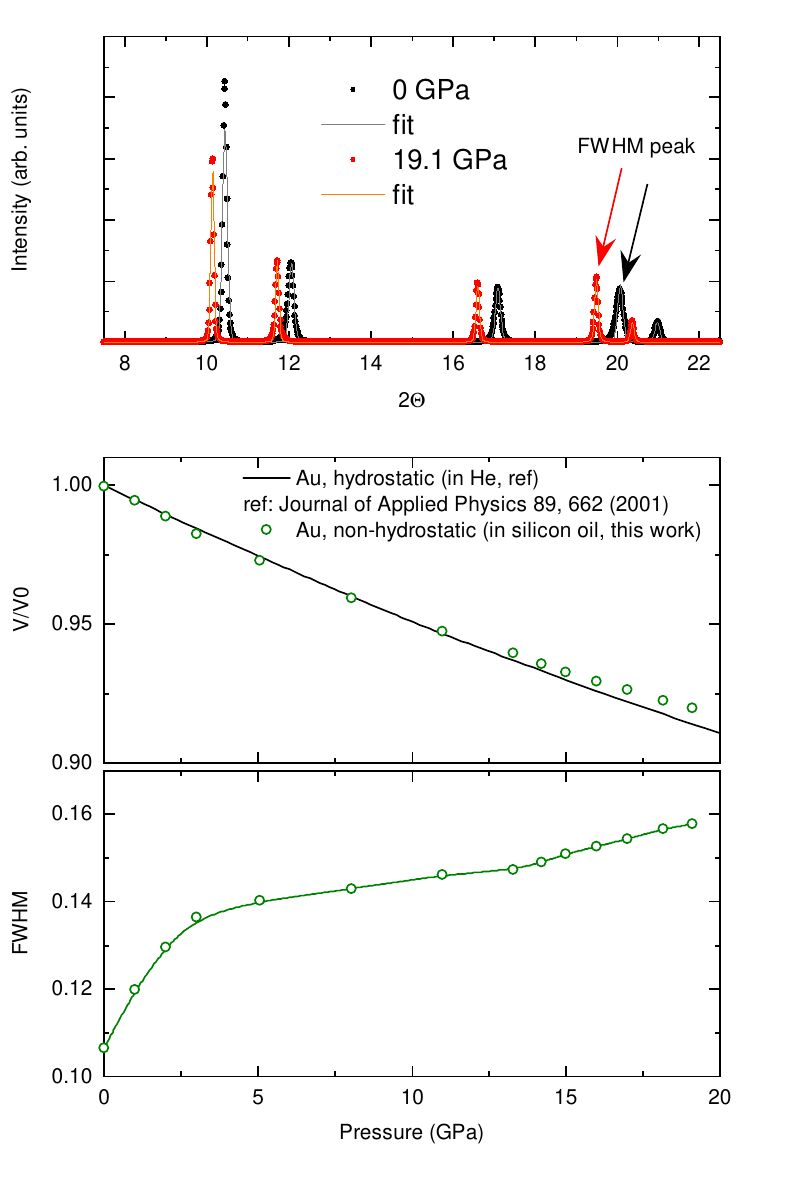}
\caption{\label{fig:gold}
Top: powder patterns of the gold reference taken at 0\,GPa and 19.1\,GPa with silicon oil as pressure-transmitting medium. Middle: unit-cell volume of gold determined using the position of the 222 reflection (shown with the arrows) is compared to the evolution of gold under quasi-hydrostatic conditions~\cite{takemura2001}. Bottom: full-width at half-maximum of the 222 reflection increases with pressure and indicates the growing nonhydrostaticity above 12\,GPa.
}
\end{figure}

\begin{figure}
\includegraphics[width=12cm]{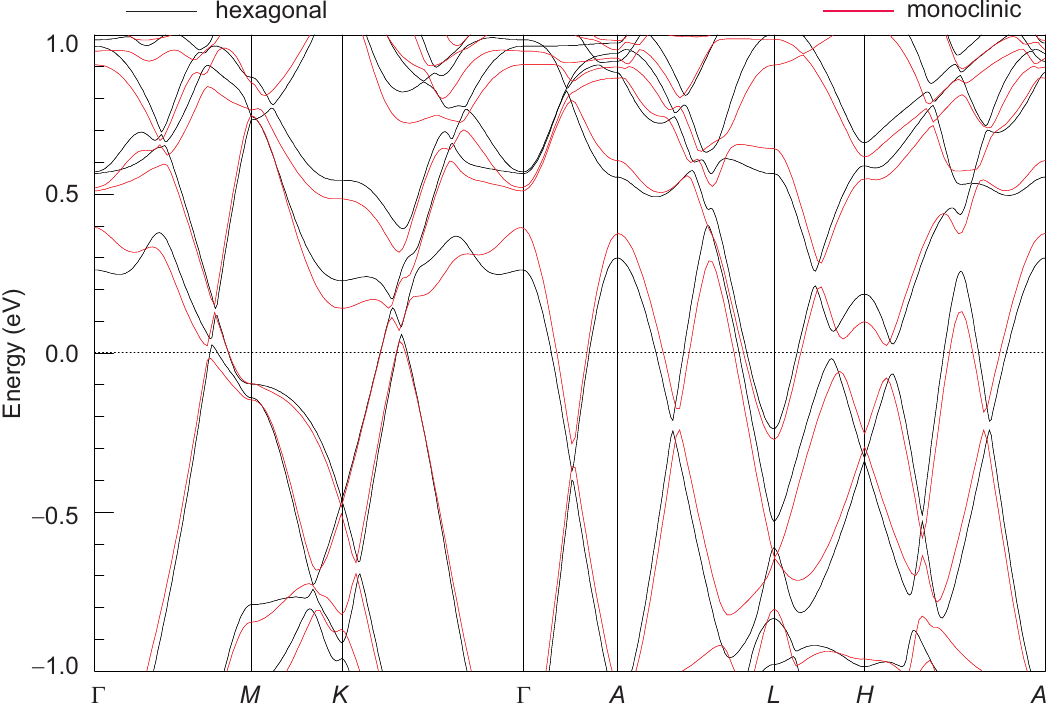}
\caption{\label{fig:bands-SM}
Direct comparison of the band structures calculated for the hexagonal and monoclinic phases of CsV$_3$Sb$_5$ at 18.2\,GPa. Atomic positions of the hexagonal structure from Ref.~\cite{tsirlin2022} have been used. The $k$-path of the hexagonal structure is chosen.
}
\end{figure}
\end{widetext}

\end{document}